\documentclass{article}
\usepackage{lajolla2006,graphicx}
\frompage{000} \topage{000}                                              
\def\rightmark{Di-hadron and Tri-hadron correlation and Mach-like cone structure }
\def\leftmark{Y. G. Ma et al.}

\title{Di-hadron and Tri-hadron correlation and Mach-like cone structure in AMPT model}

\authors{
{Y. G. Ma$^{a}$, G.L. Ma$^{a,b}$, S. Zhang$^{a,b}$, X.Z.
Cai$^{a}$, J.H. Chen$^{a,b}$, Z.J. He$^a$,  H.Z. Huang$^c$, J.L.
Long$^a$, W.Q. Shen$^{a}$, X.H. Shi$^{a,b}$,
J.X. Zuo$^{a,b}$}\\[2.812mm]
{\normalsize \hspace*{-8pt} $^a$ Shanghai Institute of Applied Physics, Chinese Academy of
Sciences, Shanghai 201800, China, \\
\hspace*{-8pt} $^b$ Graduate School of the Chinese Academy of Sciences, Beijing 100080, China, \\
\hspace*{-8pt} $^c$ University of California, Los Angeles,CA90095, USA
 } }

\abstract{In a framework of a multi-phase transport model with
both partonic and hadronic interactions, azimuthal correlations
between trigger particles and associated scattering particles in
Au + Au collisions at $\sqrt{s_{NN}}$ = 200 GeV/$c$ have been
studied by the mixing-event technique. The Mach-like structure has
been observed in correlation function for central collisions. It
is shown that the Mach-like structure is basically born in the
partonic process and further developed in hadronic rescattering
process. However, hadronic rescattering alone cannot reproduce the
amplitude of Mach-like cone on away side, therefore partonic
cascade process is necessary to describe the amplitude of
Mach-like cone on away side in experiment. In addition,
three-particle correlations have been investigated in central Au +
Au collisions with the AMPT model, and the results support the
conclusion that partonic cascade processes enhance the opening
angle of Mach-like cone structures. }


\begin{document}

\maketitle
\setcounter{page}{1}

\section{Introduction}

The strong suppression of high-$p_{T}$ particle
yield~\cite{hiptsuppress} and the disappearance of one jet in
back-to-back jet correlation~\cite{hard-hard-ex} have been
observed in Au + Au central collisions at $\sqrt{s_{NN}}$= 200
GeV/c, which can be interpreted by jet quenching
mechanism~\cite{HIJING}. On the other hand, the loss energy will
be redistributed in the soft $p_T$ region
\cite{soft-soft-th1,soft-soft-th2,soft-soft-th3,soft-soft-th4}.These
soft associated particles which carry the loss energy have been
reconstructed via two-particle angular correlation of charged
particles in STAR experiment\cite{soft-soft-ex}, which will
constrain models for the description of production mechanisms of
high $p_{T}$ particles, and may shed light on the underlying
energy loss mechanisms and the degree of equilibration of jet
products in the medium.

The interesting Mach-like structure has been observed recently in
two-particle correlation function in Au + Au collisions at
$\sqrt{s_{NN}}$ = 200
GeV/c~\cite{sideward-peak1,sideward-peak2,sideward-peak3}.
Corresponding theoretical studies have just started with many new
ideas. For instance, it was proposed that a Mach shock wave will
happen when the jet travels faster than sound in the medium
\cite{Stocker,Casalderrey1,Ruppert,Thorsten,Chaudhuri}; the
Mach-like structure can also be produced with a Cherenkov
radiation model \cite{Koch}; it was attributed to medium dragging
effect in Ref.~\cite{Armesto}.

In this work, we shall study two-particle and three-particle
correlations between trigger particle and associated particle(s)
and investigate strange Mach-like structure by using a dynamical
transport model: a multi-phase transport model ( AMPT )
\cite{AMPT}. We applied mixing-event technique to AMPT results as
people made the analysis for RHIC data and reproduced two-particle
and three-particle correlations with AMPT model. It is found that
both parton cascade and hadronic rescattering can produce the
apparent associated particle correlation as well as Mach-like
structure. But the pure hadronic rescattering mechanism can not
reproduce the amplitude of Mach-like cone on away side, therefore
parton cascade process seems indispensable.

\section{Model Introduction}
AMPT model \cite{AMPT}  is a hybrid model  which consists of four
main components: the initial condition, partonic interactions, the
conversion from partonic matter into hadronic matter and hadronic
interactions. The initial condition, which includes the spatial
and momentum distributions of minijet partons and soft string
excitation, are obtained from the HIJING model \cite{HIJING}.
Excitation of strings will melt strings into partons. Scatterings
among partons are modelled by Zhang's parton cascade model  (ZPC)
\cite{ZPC}, which at present includes only two-body scattering
with cross section obtained from the pQCD with screening mass. In
the default AMPT model \cite{DAMPT} partons are recombined with
their parent strings when they stop interaction, and the resulting
strings are converted to hadrons by using a Lund string
fragmentation model \cite{Lund}. In the AMPT model with string
melting \cite{SAMPT}, a simple quark coalescence model is used to
combine partons into hadrons. Dynamics of the subsequent hadronic
matter is then described by A Relativistic Transport (ART) model
\cite{ART}. Details of the AMPT model can be found in a recent
review \cite{AMPT}. It has been shown that in previous studies
\cite{SAMPT} the partonic effect could not be neglected and a
string melting AMPT is much more appropriate than the default AMPT
when the energy density is much higher than the critical density
for the pQCD phase transition \cite{AMPT,SAMPT,Jinhui}. In the
present work, the parton interaction cross section in AMPT model
with string melting is 10mb.

\section{Two-particle Correlation Analysis}

\subsection{Analysis Method}

\begin{figure}[htb]
\centerline{\includegraphics[width=8cm]{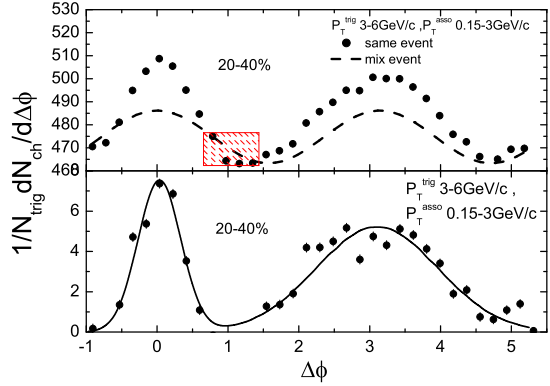}}
\caption{\label{fig:2pmethod}(a): The associated hadron
$\Delta\phi$ distribution  for the trigger hadrons with
$3<p_{T}^{trig}<6$ GeV/$c$ and the associated hadrons with
$0.15<p_{T}^{assoc}<3$ GeV/$c$ (circles) where the background
(dash line) is not subtracted for 200 GeV/$c$ Au + Au collisions
at 20-40\% centrality within AMPT model. The dash area is the
region of ZYAM normalization (see texts for detail); (b): The
associated hadron $\Delta\phi$ distribution where the background
has been subtracted by mixing-event technique, the solid line is
its two-Gaussian fit.}
\end{figure}

In order to reproduce the soft (or hard )  associated-hadron
correlations, we use the mixing-event technique in our analysis.
Two kinds of  $p_{T}$ window cuts for trigger and associated
particles are used, one is $3 < p_{T}^{trig} < 6$ GeV/$c$ and
$0.15< p_{T}^{assoc} < 3$ GeV/$c$ (we call it as "soft" associated
hadrons since the soft particles are dominated), another is $2.5 <
p_{T}^{trig} < 4$ GeV/$c$ and $1.0 < p_{T}^{assoc} < 2.5$ GeV/$c$
(we call it as "hard" associated hadrons since there are more hard
particles than the previous "soft" component). Both trigger and
associated particles are selected with pseudo-rapidity window
$|\eta| < 1.0$. In the same events, the correlation pairs of the
associated particles with trigger particles are accumulated to
obtain $\Delta\phi = \phi - \phi_{trig}$ distributions. In order
to remove the background which is expected to mainly come from the
effect of elliptic flow \cite{soft-soft-ex,sideward-peak2},
so-called mixing-event method is applied to simulate  its
background. In this method, we mixed two events which have very
close centrality into a new mixing event, and extracted
$\Delta\phi$ distribution which is regarded as the respective
background. When subtracting the background from the same events,
ZYAM (zero yield at minimum) assumption is adopted as did in
experimental analysis \cite{sideward-peak2}.
Figure~\ref{fig:2pmethod} gives us the associated hadron
$\Delta\phi$ distributions  for the trigger hadrons with $3 <
p_{T}^{trig} < 6$ GeV/$c$ and the associated hadrons with $0.15 <
p_{T}^{assoc} < 3$ GeV/$c$  before and after subtracting the
background in  200 GeV/$c$ Au + Au collisions at 20-40\%
centrality  within AMPT model.

\subsection{Results and Discussions}

\begin{figure}
\centerline{\includegraphics[width=8cm]{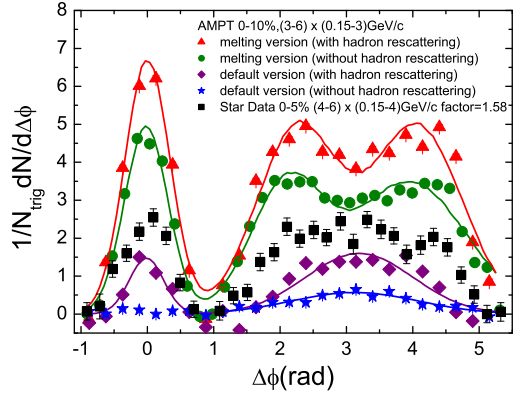}}
\caption{\label{fig:soft_machshape}Soft scattered associated
hadron $\Delta\phi$ correlations for the trigger hadrons with $3.0
< p_{T}^{trig} < 6.0 $GeV/$c$  and the associated hadrons with
$0.15 < p^{assoc}_T < 3.0 $ GeV/$c$ in most central Au + Au at
$\sqrt{s_{NN}}$ = 200 GeV/c in AMPT model. Triangles: AMPT melting
version (later we just call as melting) after hadronic
rescattering; circles: melting before hadronic rescattering;
diamonds: AMPT default version (later we just call as default)
after hadronic rescattering; stars: default before hadronic
rescattering; squares: experimental data from
Ref~\cite{soft-soft-ex} where  $4.0 < p^{trig}_{T} < 6.0 $ GeV/$c$
and $0.15 < p^{assoc}_T < 4.0$ GeV/$c$ is taken.}
\end{figure}

In order to increase the statistical amount of trigger particles
in our calculation, we set $p_{T}$ range for trigger particles to
$3 < p_{T}^{trig} < 6$ GeV/$c$ and for associated particles to
$0.15 < p_{T}^{assoc} < 3$ GeV/$c$, and pseudo-rapidity range to
$|\eta| < 1.0$ both for trigger and associated particles in our
analysis. Both trigger and associated particles are selected with
$|\eta| < 1.0$. Figure.~\ref{fig:soft_machshape} presents the soft
scattered associated hadron $\Delta\phi$ correlations in most
central Au + Au collisions at $\sqrt{s_{NN}}$=200 GeV/c  under
different conditions. In order to compare our results with
experimental data which give the correlations among associated
charged hadrons, the experimental data are multiplied by a factor
of 1.58 to account for the contribution from neutral hadrons
\cite{soft-soft-ex,factor}. From the figure,  we can see that the
hadronic rescattering increases Mach-like $\Delta\phi$
correlations not only for the melting AMPT but also for the
Default AMPT. In the melting AMPT, there are very strong Mach-like
correlations before hadronic rescattering, which indicates that
Mach-like structure has been formed in parton cascade process.

\begin{figure}
\centerline{\includegraphics[width=10cm]{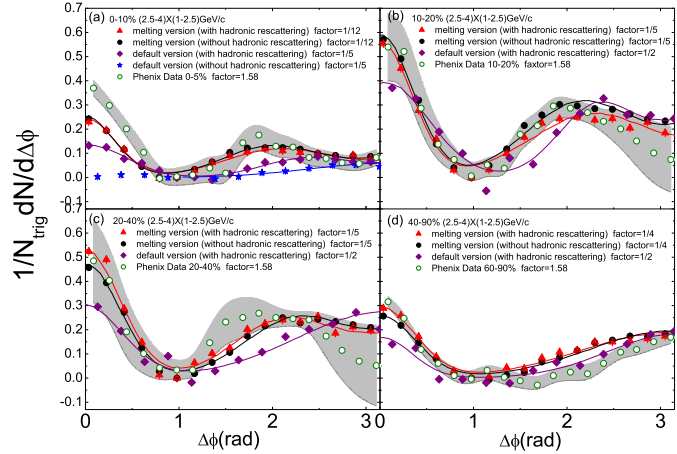}}
\caption{\label{fig:hard_machshape}Hard scattered associated
hadron $\Delta\phi$ correlations in Au + Au at 200 GeV/$c$ at
different centralities for correlated hadrons with $2.5 <
p^{trig}_T < 4.0$ GeV/$c$ and $1.0 < p_{T}^{assoc} < 2.5$ GeV/$c$
in AMPT model. Triangles: melting after hadronic rescattering;
Full circles: melting before hadronic rescattering; Diamonds:
default after hadronic rescattering; Stars: default before
hadronic rescattering; Open circles: experimental data from
Ref.~\cite{sideward-peak2}; Hatched areas give the experimental
uncertainty.}
\end{figure}

For hard scattered  associated particles ($2.5 < p^{trig}_T < 4.0
$GeV/$c$ and $1.0 < p_{T}^{assoc} < 2.5$ GeV/$c$), Figure
~\ref{fig:hard_machshape} shows $\Delta\phi$ correlations in 200
GeV/$c$ Au + Au at different centralities in different conditions.
(Note that here  our $p_{T}$ cut is the same as PHENIX cut,
But$|\eta|$ cut: $|\eta| < 1.0$ in our model is different from
$|\eta| < 0.35$ at PHENIX to increase the statistics in our
simulations.) It was found that the effect on $\Delta\phi$
correlations from hadronic rescattering is much smaller than soft
scattered associated particles, which may indicate that fewer hard
associated hadrons are from hadronic rescattering. In addition,
Mach-like structures are observed on away sides in both melting
AMPT and default AMPT. However we should point out  that in the
default AMPT the Mach-like structures can only be observed after
hadronic rescattering. For the comparison with the data, the
Melting AMPT gives more reasonable splitting between two Mach-like
peak in away side. It indicates that partonic interaction is more
important to describe the Mach-like structure.

\section{Three-particle Correlation Analysis}
\subsection{Analysis Method}

\begin{figure}[htb]
\begin{center}
\hfill
    \begin{minipage}[t]{0.47\textwidth}
        \includegraphics[width=1.00\textwidth]{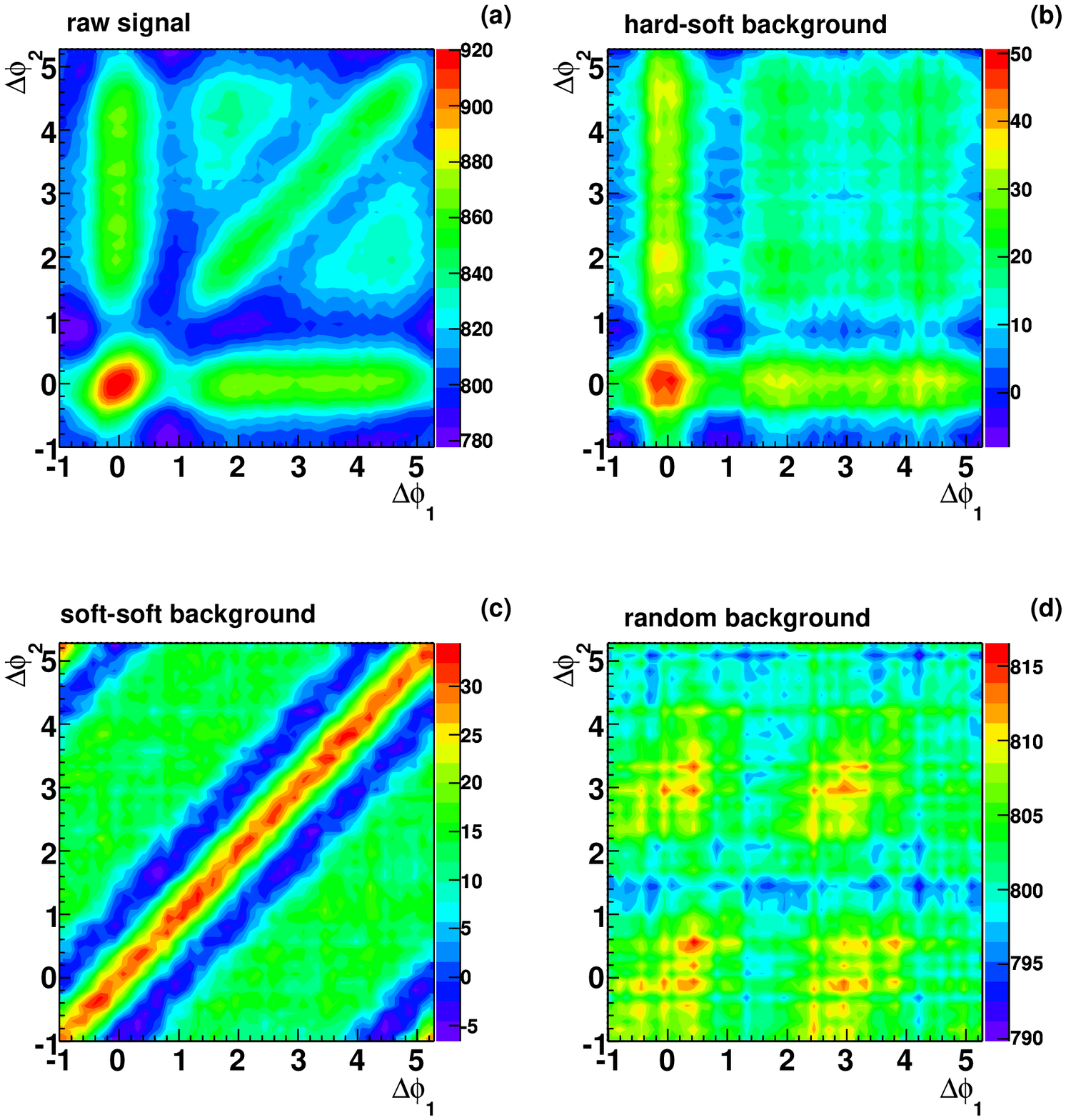}
        \vspace*{-1cm}
        \caption{ Three-particle
           correlations in the 10\% most central 200GeV/c Au + Au collisions (with string melting and  hadron
           rescattering scenario). (a) Raw signal. (b) Hard-soft background. (c) Soft-soft background. (d) Random background.}
            \label{fig:3p_process}
        \end{minipage}
        \hfill
        \begin{minipage}[t]{.47\textwidth}
        \includegraphics[width=1.00\textwidth]{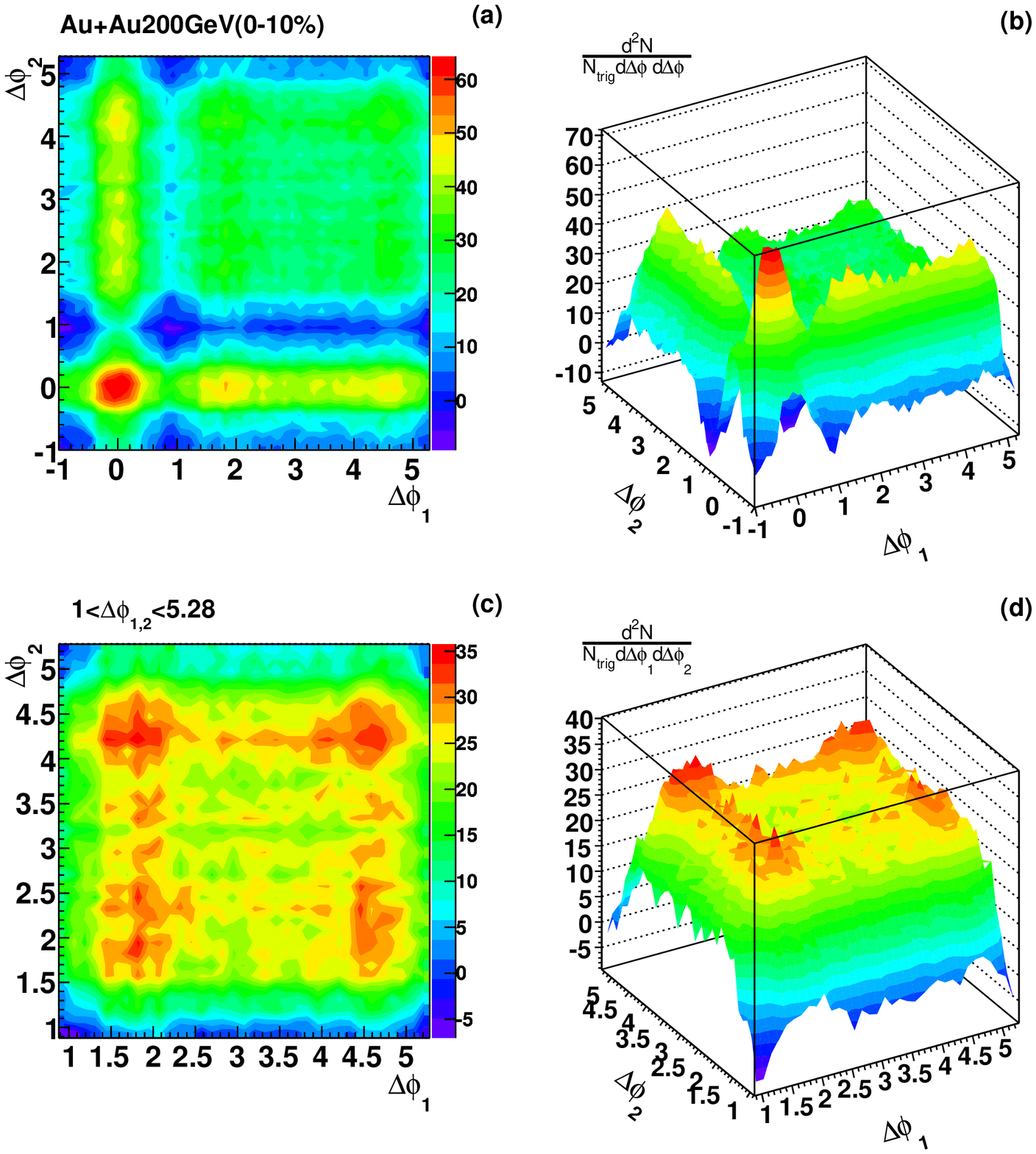}
            \vspace*{-1cm}
        \caption{Background
subtracted 3-particle correlations in the 10\% most central 200
GeV/c Au + Au collisions (with string melting and after hadron
rescattering). (a) and (b): Background subtracted 3-particle
correlations ($-1<\Delta\phi_{1,2}<5.28$). (c) and (d): Segmental
Background subtracted 3-particle correlations
($1<\Delta\phi_{1,2}<5.28$).}
    \label{fig:3p_signalAuAu}
    \end{minipage}
    \end{center}
\end{figure}

The mixing-event technique is used in our three-particle
correlation analysis. Only hard $p_{T}$ window cut for trigger and
associated particles is selected as  $2.5 < p_{T}^{trig} < 4$
GeV/$c$ and $1.0 < p_{T}^{assoc} < 2.5$ GeV/$c$ in the analysis.
Both trigger and associated particles are selected with
pseudo-rapidity window $|\eta| < 1.0$. In the same events, raw
3-particle correlations in $\Delta\phi_{1} = \phi_{1} -
\phi_{trig}$and $\Delta\phi_{2} = \phi_{2} - \phi_{trig}$ are
accumulated. The figure\ref{fig:3p_process}(a) show raw 3-particle
correlations in the 10\% most central Au+Au at 200GeV/$c$
collisions with string melting and after hadron rescattering.
Three background contributions are expected in raw signal. The
first one is a trigger-associated pair combined with a background
associated particle, which was reproduce by mixing
trigger-associated pairs with another false associated particle
that from another events (Figure \ref{fig:3p_process}(b)). We call
it hard-soft background. The second one is a associated particle
pair combined with a background trigger particle, which was
reproduce by mixing associated particle pairs with another false
trigger particle that from another events (Figure
\ref{fig:3p_process}(c)). We call it soft-soft background. The
last one is a random background, which are produce by mixing
trigger particle and two associated particles respectively from
three different events (Figure \ref{fig:3p_process}(d)). When
subtracting the background from the same events, we normalize the
strip of $0.8<|\Delta\phi_{1,2}|<1.2$ to zero.
Figure~\ref{fig:3p_signalAuAu} (a) and (b) give background
subtracted three-particle correlations in the 10\% most central
200GeV/c Au+Au collisions with string melting and hadronic
rescattering mechanism. In order to see the 3-particle
correlations among trigger particle and two away-side associated
particles on away side, the zoom-in segmental 3-particle
correlations ($1<\Delta\phi_{1,2}<5.28$) are shown in
figure~\ref{fig:3p_signalAuAu} (c) and (d).

\subsection{Results and Discussions}

In panel (c) of figure~\ref{fig:3p_signalAuAu}, three interesting
regions will be investigated. The first one is 'center' region
($|\Delta\phi_{1,2}-\pi|<0.8$) where three-particle correlations
mainly come from trigger particle and two associated particles in
the center of away side. The second one is 'deflected'
region($|\Delta\phi_{1,2}-(\pi\pm1)|<0.8$) where three-particle
correlations reflect three-particle correlations among trigger
particle and two associated particle in one same cone of away side
and are expected to be due to the sum of away-side jets deflected
by radial flow. The third one is 'cone' region
($|\Delta\phi_{1}-(\pi\pm1)|<0.8$ and
$|\Delta\phi_{2}-(\pi\mp1)|<0.8$) where it gives three-particle
correlations among trigger particle and two associated particles
from two different cones of away side. It was predicted that
'cone' correlations may be caused by Mach shockwave effect.

It was observed that all 'center', 'deflected' and 'cone'
three-particle correlations exist in most central Au+Au collisions
at $\sqrt{s_{NN}}$ = 200 GeV/c (0-10\% centrality) under melting
AMPT version after hadronic rescattering. It indicates that these
three different mechanisms may contribute to three-particle
correlation in most central Au+Au collisions, i.e. one part of
associated particles go through reaction system , one part of
associated particles are deflected, and other part of associated
particles produce Mach-like correlations.

\section{Summary}

In summary, the origin of the Mach-like correlation for soft or
hard scattered associated particles was investigated in the
framework of the AMPT model which includes two dynamical
processes, namely parton cascade and hadronic rescattering.  By
comparing the different two-particle and three-particle
correlation calculation results before or after hadronic
rescattering, with or without string melting mechanism, it is
argued that the associated particle correlation and Mach-like
structure have been formed before hadronic rescattering, which
indicates that these kinds of correlations are born in the
partonic process and are further developed in later-on hadronic
rescattering process. In our work, it is shown that hadron
rescattering mechanism can produce the associated particle
correlation, but it can not give big enough splitting for
away-side Mach-like peaks in confronting with experimental data.
In this context, the parton cascade mechanism is essential for
describing experimental Mach-like structure.

This work was supported in part by the Shanghai Development
Foundation for Science and Technology under Grant Numbers
05XD14021, the National Natural Science Foundation of China under
Grant No 10328259, 10135030, 10535010.

\end{document}